\documentclass{elsart1p}

 \usepackage{epsfig}
\usepackage{subfigure}
\usepackage{rotating}

\usepackage{amssymb}
\usepackage{amsmath}

\newcommand{\trace}{\text{Tr} }

\newcommand{\PiFrho}[3]{\Pi^{\; #1}_{(#2) \, #3}}

\newcommand{\intmeasure}[2]{\frac{d^{#1} #2}{(2 \pi)^{#1}}}

\newcommand{\myalpha}{\kappa}

\begin{document}

\begin{frontmatter}



\title{Turbulent spectra in real-time gauge field evolution\thanksref{talk}}

\thanks[talk]{Talk presented at Strong and Electroweak Matter 2008 (Amsterdam, August 2008)}

\author{Sebastian Scheffler}
\ead{sebastian.scheffler@physik.tu-darmstadt.de}

\address{Institut f\"ur Kernphysik, Technische Universit\"at Darmstadt, Schlossgartenstr. 9, 64285 Darmstadt, Germany}

\begin{abstract}
We investigate ultraviolet fixed points in the real-time evolution of non-Abelian gauge fields. Classical-statistical lattice simulations reveal equal-time correlation functions with a spectral index $\myalpha = 1.5$. Analytical understanding of this result is achieved by employing a 2PI- loop expansion for the quantum theory. 
\end{abstract}

\end{frontmatter}

\section{Introduction}
\label{section-intro}

The occurrence of turbulent power law spectra after the saturation of chromo-Weibel instabilities has been established by several authors~\cite{power-laws}. However, the value of the spectral index characterising these spectra is still under debate. \\

We investigate this question in two ways~\cite{our-letter}: Firstly, we carry out a perturbative, analytical calculation for the UV- part of the spectrum similarly as in~\cite{Berges:2008wm}. This yields a spectral index $\myalpha = 1.5$ in the equal-time correlation function of the gauge fields. Subsequently, we run classical-statistical numerical simulations starting from nonequilibrium initial conditions. We find that the spectral index from the numerical solution agrees with the result from the analytical computation.

\section{Analytics}
\label{section-analytics}

Our interest in this work focuses on the anti-commutator expectation value
 \begin{equation}\label{def-F-quantum}
F_{\mu \nu}^{ab}(x,y) \equiv \frac{1}{2} \langle \, \{ A^a_{\mu}(x), A^b_{\nu}(y) \} \, \rangle
\end{equation}
of the gauge field. It appears in a decomposition of the full propagator according to
\begin{equation}
 G^{ab}_{\mu \nu}(x, y) \equiv F^{ab}_{\mu \nu}(x,y) - \frac{i}{2} \rho^{ab}_{\mu \nu}(x,y) \, \text{sign}( x^0 - y^0 ) 
\end{equation}
where $\rho^{ab}_{\mu \nu}(x,y) \equiv i \langle \, [ A^a_{\mu}(x), A^b_{\nu}(y) ] \, \rangle $ is the spectral function. The nonlocal part of the self-energy can be decomposed in a similar way as the propagator~\cite{PRD2004},
\begin{equation}\label{def-PiFrho}
\Pi_{\mu \nu}^{a b}(x, y) \equiv \PiFrho{a b}{F}{\mu \nu}(x, y)  - \frac{i}{2} \PiFrho{a b}{\rho}{\mu \nu}(x, y) \, \text{sign}( x^0 - y^0 ) \; ,	
\end{equation}
with (anti-)symmetric $\PiFrho{}{F}{} $ ( $\PiFrho{}{\rho}{}$). All these objects are diagonal in the colour indices and thus we will drop the latter in what follows.

The goal of this work is to study fixed-point solutions. These are invariant under time translations and we additionally assume  spatial translation invariance as well. This implies that all correlation functions and self-energies only depend on the displacement $x - y$, that is $F_{\mu \nu}(x, y ) \equiv F_{\mu \nu}(x - y) \equiv F_{\mu \nu}(x^0 - y^0, \vec{x} - \vec{y}) $, and analogously for the other objects. A solution with these properties satisfies the following identity in Fourier space~\cite{Berges:2008sr}:
 \begin{equation}\label{stationarity-condition}
  \PiFrho{\mu \gamma}{\rho}{} ( p ) F_{\gamma \nu}(p) - \PiFrho{\mu \gamma }{F}{}(p) \rho_{\gamma \nu}( p) = 0  \; .
\end{equation}
In order to find an analytical solution we will later integrate this equation over $\vec{p}$.

Now we make the following scaling ansatz for the symmetric and anti-symmetric correlation functions:
\begin{equation}\label{scaling-assumption}
F_{\mu \nu}(s p ) = s^{- (2 + \myalpha )} F_{\mu \nu}(p) \; \;  \text{  and  } \; \; \rho_{\mu \nu}(s p ) = s^{-2} \rho_{\mu \nu}(p)  \; .
\end{equation}
The partial Fourier transform of such a solution for $F$ possesses the property
\begin{equation}\label{F-scaling-expectation}
F_{\mu \nu}(\Delta t = 0, s \vec{p} ) = s^{- (1 + \myalpha )} F_{\mu \nu}(\Delta t =0, \vec{p}) \; .
\end{equation}

\begin{figure}[t!]
\begin{center}
\epsfig{file=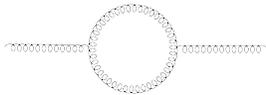, width=3.5cm, angle =0 }
\caption{The leading order contribution to the gluon self-energy in a perturbative expansion. The lines represent full propagators.}
\label{fig-loop-diagram} 
\end{center}
\end{figure}

In a perturbative calculation the dominant contribution to the self-energies in the UV stems from the diagram depicted in Fig.~\ref{fig-loop-diagram}. Using the perturbative three- vertex for the SU(2)- gauge theory we can write the self-energies as
\begin{equation}\label{self-energies-2}
\begin{split}
\PiFrho{\mu \nu}{F}{}(p) & = - g^2 \int \intmeasure{4}{k} \int \intmeasure{4}{q} \,  \delta(p + k + q ) \, \bigl\{ \, F_{\kappa \rho}(k) F_{\lambda \sigma}(q) - \rho_{\kappa \rho}(k) \rho_{\lambda \sigma}(q) \, \bigr\} \\
& \hspace{-1.2cm} \times \bigl[ \, g^{\mu \kappa}(p - k)^{\lambda} +  g^{\kappa \lambda} (k -q)^{\mu} + g^{\lambda \mu } (q - p)^{\kappa} \, \bigr] \bigl[ \, g^{\nu \rho}(p-k)^{\sigma} + g^{\rho \sigma}(k-q)^{\nu} + g^{\sigma \nu} (q-p)^{\rho}  \, \bigr] \\
\PiFrho{\mu \nu}{\rho}{}(p) & = + g^2 \int \intmeasure{4}{k}  \int \intmeasure{4}{q} \, \delta(p + k + q ) \, \bigl\{ \, F_{\kappa \rho}(k) \rho_{\lambda \sigma}(q) + \rho_{\kappa \rho}(k) F_{\lambda \sigma}(q) \, \bigr\} \\
& \hspace{-1.2cm} \times \bigl[ \, g^{\mu \kappa}(p - k)^{\lambda} +  g^{\kappa \lambda} (k -q)^{\mu} + g^{\lambda \mu } (q - p)^{\kappa} \, \bigr] \bigl[ \, g^{\nu \rho}(p-k)^{\sigma} + g^{\rho \sigma}(k-q)^{\nu} + g^{\sigma \nu} (q-p)^{\rho}  \, \bigr] \; .
\end{split}
\end{equation}
In the classical limit where $F^2 \gg \rho^2$ the contribution to $\PiFrho{\mu \nu}{F}{}$ of the two spectral functions $\rho$ in the curly brackets vanishes. Inserting the above expressions for the self-energies into Eq.~\eqref{stationarity-condition}, integrating over the spatial $\vec{p}$ and carrying out a generalised Zakharov transformation similarly as in~\cite{Berges:2008wm} we obtain 
\begin{equation}\label{result-alpha}
 \myalpha = \frac{3}{2} \; .
\end{equation}
Further details of the calculation can be found in~\cite{our-letter}. We emphasise that without the classicality assumption there would not be a fixed-point solution.

\section{Numerical results}
\label{section-numerics}

Our numerical setup is the same as in~\cite{Berges:2007re}. We briefly summarise the details that are relevant for the present discussion.

We use a standard lattice discretisation scheme in terms of link variables $U_{x, \mu}$ representing parallel transporters from the lattice site $x$ to $x + \hat{\mu} $. Given the link variables one can compute the gauge fields as $A_{\mu}^b(t, \vec{x}) \simeq - i \, \trace \, \bigl( \sigma^b U_{\mu}(t, \vec{x}) \bigr) / (2 a g) $ where $\sigma^b \, , \, b = 1, 2, 3$, are the Pauli matrices and $a $ is the lattice spacing.

We compute the real time evolution of gauge fields by solving the discrete equations of motion derived from a variation of the Wilson lattice action in Minkowski spacetime. The initial conditions are sampled from a Gaussian distribution such that $ \langle \, \vert \, A^b_j(t_0,\vec{p}) \, \vert^2 \, \rangle \, \propto \exp \{ - 0.5 ( p_x^2 + p_y^2 )/ \Delta^2 - 0.5 (p_z^2) / \Delta_z^2 \} $ for the gauge fields in momentum space. The expectation values occurring in the definitions of the correlation functions~\eqref{def-F-quantum} have to be taken with respect to this distribution.

\begin{figure}[t!]
\begin{center}
\hspace{0.5cm}
\epsfig{file=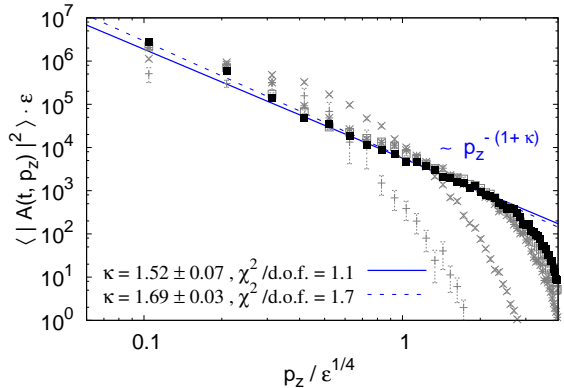, width=5.5cm, angle=270}
\caption{Spectra of $F_{\mu \nu}(\Delta t =0, \vec{x}) = \langle \, A_{\mu}(t, \vec{x}) A_{\nu}(t, 0 )  \, \rangle $ , $ \mu = \nu$ . Grey symbols: Spectra at times $ t = 212 - 705 \, \epsilon^{-1/4}$. Black squares: $t = 940 \, \epsilon^{-1/4} $. The blue lines represent the fit results, confer the text for details. The momenta are perpendicular to the spatial direction of the gauge fields.}
\label{fig-power-laws} 
\end{center}
\end{figure}

We numerically follow the time evolution of the gauge fields to times far beyond the saturation of the instabilities. Computing the Fourier spectrum\footnote{Precisely speaking, $F_{\mu \nu}$ here denotes the classical approximation to the object defined in Eq.~\eqref{def-F-quantum}.} of $F_{\mu \nu}(\Delta t =0, \vec{x}) = \langle \, A_{\mu}(t, \vec{x}) A_{\nu}(t, 0 )  \, \rangle $  at late times in Coulomb gauge\footnote{Our gauge fixing employs the stochastic overrelaxation algorithm described in~\cite{Cucchieri:1995pn}. } we observe in Fig.~\ref{fig-power-laws} that the spectrum converges towards a stationary curve. Fitting a power law ansatz $ F_{\mu \nu}(0, \vec{p}) \overset{!}{=} C | \vec{p} |^{-(1+\myalpha)} $ as in Eq.~\eqref{F-scaling-expectation} to the latest available spectrum on the domain $ 0.6 \,  < p \cdot \epsilon^{-1/4} < \, 2.0 $ we obtain
\begin{equation}\label{result-alpha-numerical}
\myalpha = 1.52 \pm 0.07 \; .
\end{equation}
We can produce a slightly higher exponent of $\myalpha = 1.69 \pm 0.03 $ by including also the lowest momenta into the fit. However, this increases $\chi^2$ per degree of freedom to 1.7, in contrast to $\chi^2 / \text{  d. o. f. } = 1.1$ for the fit underlying~\eqref{result-alpha-numerical}. Therefore, the result quoted in~\eqref{result-alpha-numerical} is more reliable and we note that it is in excellent agreement with Eq.~\eqref{result-alpha}. In any case our data is incompatible with $\myalpha =2$ and also with the thermal value $\myalpha =1 $. The fit results are visualised in Fig.~\ref{fig-power-laws}.

\section{Conclusions}
\label{section-conclusions}
The analytical and numerical results for the spectral index are in excellent agreement. Their numerical value also agrees with the one found for the UV- spectrum of scalar field theories in the context of early-universe cosmology~\cite{Berges:2008wm,Micha:2004bv}.  

What is the meaning of these findings? Comparing the present results to~\cite{Berges:2008wm,Micha:2004bv} shows that some features of QCD are quantitatively the same as for scalar theories. This might allow to establish the notion of universality far from equilibrium. In fact, quantum corrections will eventually drive the system away from the classical fixed-point solution in the UV. However, they will not alter the infrared behaviour, as it is known from scalar theories~\cite{Berges:2008wm}. It remains as a future project to search for nonthermal fixed points in the IR- sector of the gauge theory. Also, it would be desirable to discuss fixed points in terms of gauge-invariant quantities. 

I thank my collaborators J\"urgen Berges and D\'enes Sexty for their contributions to this project.



\end{document}